\documentclass[preprint]{revtex4}
\usepackage{amsmath,amssymb,dsfont}
\usepackage{epsfig,graphics}

\newcommand{\smfrac}[2]{{\textstyle\frac{#1}{#2}}}
\renewcommand{\t}{\omega t}

\begin{document}
\title{Analytical Approach to Wave Packet Dynamics\\ of Laser-Driven Particles beyond the Dipole Approximation}
\author{M. Verschl}\email{mario.verschl@mpi-hd.mpg.de}
\author{C.H. Keitel}\email{keitel@mpi-hd.mpg.de}
\affiliation{Max-Planck-Institut f\"ur Kernphysik, Saupfercheckweg 1, 69117 Heidelberg, Germany}
\affiliation{Theoretische Quantendynamik, Physikalisches Institut, Universit\"at Freiburg,\\ Hermann-Herder-Str. 3, 79104 Freiburg, Germany}

\begin{abstract}
An analytical approach to quantum mechanical wave packet dynamics of laser-driven particles is presented. The time-dependent Schr\"odinger equation is solved for an electron exposed to a linearly polarized plane wave of arbitrary shape. The calculation goes beyond the dipole approximation, such that magnetic field effects like wave packet shearing are included. Analytical expressions for the time-dependent widths of the wave packet and its orientation are established. These allow for a simple understanding of the wave packet dynamics. 
\end{abstract}

\maketitle

\newpage
\section{Introduction}
Experimental progress in laser physics has enabled to reach intensities that can accelerate electrons to velocities comparable to the speed of light. The theoretical description, i.e. the solutions to the Dirac equation describing particles that are exposed to plane light waves (Volkov states) have long been known \cite{Volkov}. Since a particle that occupies such a Volkov state is not localized in space, the dynamics is neither very intuitive nor does it describe a realistic situation.
Different methods have been developed to describe localized wave packets. One approach is to numerically solve the Dirac equation as in \cite{Joachain,Knight,Grobe,Mocken}. A further possibility is to superimpose Volkov solutions to find the dynamics of wave packets \cite{Roso}. It has been found that the wave packets deform and spread non-isotropically for high laser intensities. Although this behavior is seen when the wave packets are plotted, there is no detailed understanding of their dynamics. 

Going to lower intensities where the particle dynamics can be described by means of the Schr\"odinger equation the dipole approximation can often be employed. The consequence is that the magnetic field is neglected. This case can easily be handled analytically (see for example \cite{Friedrich}), but without the magnetic field the deformation effects do not appear and the wave packet spreading reduces to that of a free particle.

The purpose of this paper is to establish the wave packet dynamics in the regime where the dipole approximation is not valid anymore and the velocity of the particle is still slow compared to the speed of light, though non-negligible. That means, the particle can be described by the Schr\"odinger equation which includes all relativistic effects of order $1/c$. Since the magnetic field plays an important role in the regime considered here, deformation effects are to be expected. The wave packet will be characterized by the trajectory of its maximum, the principal axes and its widths. The establishment of the corresponding analytical expressions describing the wave packet will give intuitive insight into its dynamics.

Having defined the laser fields and the appropriate field expansion in the first section, the dynamics will be solved for a classical particle. Later on, this will be used for a comparison with the trajectory of the quantum mechanical wave packet. In section \ref{QM1} to \ref{QM3} the quantum mechanical wave packet dynamics will be deduced. After solving the Schr\"odinger equation for a spacially restricted area, a gaussian wave packet will be constructed. In the last section the results will be applied to a particular example of a short laser pulse.

\section{Laser Field Expansion}
The following calculation will refer to the dynamics of an electron, that is exposed to a laser pulse. The emitted light is supposed to be a linearly polarized plane wave with arbitrary shape. If the pulse travels in z-direction, the vector potential is given by
\begin{equation}
\label{LFE1}
\vec{A}=\hat{x}\cdot A(\omega t-kz)\equiv\hat{x}\cdot A(\varphi),
\end{equation}
where the polarization is in x-direction. The wave vector $k$ and the angular frequency $\omega$ are as usual connected by the speed of light $c=\omega/k$.
With the assumption that the particle be close to the plane $z=0$, i.e. $(kz)^2\ll 1$ the vector potential can be expanded:
\begin{equation}
\label{LFE2}
\vec{A}\approx\hat{x}\big[A(\omega t)-kz\cdot A'(\omega t)\big]\,.
\end{equation}
The prime denotes the derivative with respect to the phase $\varphi$, whereas a dot will be used for time derivatives. 
The following example is a useful check for the validity of the approximation: If the laser emits visible light of wave length $\lambda=800nm$, then the particle can be displaced up to a few hundred atomic units away from the plane $z=0$ without $kz$ exceeding the order of $10^{-1}$.

As in the expansion (\ref{LFE2}), the following calculations will be carried out to the first order in $(kz)$.  That means effects that are due to the z-dependence of the vector potential will be found which are not contained in the dipole approximation, where any z-dependence is eliminated. 

The electric and magnetic fields are derived from the vector potential:
\begin{subequations}
\label{LFE3}
\begin{align}
\vec{E}&=-\frac{1}{c}\dot{\vec{A}}=-\frac{\omega}{c}(A'-kz\cdot A'')\cdot\hat{x}\\
\vec{B}&=\vec{\nabla}\times\vec{A}=-k\cdot A'\cdot\hat{y}\,.
\end{align}
\end{subequations}
As opposed to the dipole approximation, the magnetic field is not equal to zero here.

\section{Classical Equations of Motion}\label{ClEq}
Although the fully relativistic dynamics of a classical particle driven by a plane wave, is well-known (see for example \cite{Salamin}), a non-relativistic treatment beyond the dipole approximation will be presented. The purpose is to find a result that can be used for a comparison of the trajectory of a quantum mechanical wave packet and the corresponding classical solution.

In the following, atomic units will be used, i.e. Planck's constant $\hbar$, the electron mass $m$ and the unit charge $e$ will be set equal to one.

The motion of a classical, non-relativistic charged particle is given by the equation
\begin{equation}
\label{ClEq1}
\ddot{\vec{x}}=\vec{E}+\frac{\dot{\vec{x}}}{c}\times\vec{B}\,.
\end{equation}
Applied to the fields (\ref{LFE3}) one arrives at the following component form:
\begin{subequations}
\label{ClEq2}
\begin{align}
\label{ClEq2a}
\ddot{x}&=-\frac{\omega}{c}(A'-kz\cdot A'')+\frac{k}{c}\cdot\dot{z}\cdot A'\\
\label{ClEq2b}
\ddot{y}&=0\\
\label{ClEq2c}
\ddot{z}&=-\frac{\omega}{c^2}\cdot\dot{x}\cdot A'
\end{align}
\end{subequations}
Using $\omega\cdot A'=\dot{A}$ equation (\ref{ClEq2a}) can be integrated and inserted into (\ref{ClEq2c}):
\begin{subequations}
\label{ClEq3}
\begin{align}
\label{ClEq3a}
\dot{x}&=-\frac{1}{c}A+\frac{\omega}{c^2}\cdot z\cdot A'+p_x\\
\label{ClEq3c}
\ddot{z}&=\frac{\omega}{c^2}\big(\frac{1}{c}A-kz\frac{1}{c}A'-p_x\big)A'
\end{align}
\end{subequations}
Here the initial conditions are such that $\dot{\vec{x}}(t=0)=\vec{p}$ and $\vec{x}(t=0)=\vec{x}_0$. Now, equation (\ref{ClEq3c}) is formally integrated and one finds the following integral equation:
\begin{equation}
\label{ClEq4}
\frac{\dot{z}}{c}=\frac{1}{2c^4}A^2-\frac{p_x}{c^3}A+\frac{p_z}{c}-\frac{\omega}{c^4}\int(kz)A'\,^2 dt
\end{equation}

At this point, some simplifications can be made. $A/c^2$ and $p_x/c$ have to be a small number such that the particle dynamics is non-relativistic. This is seen from (\ref{ClEq3a}) ($\dot{x}/c=-A/c^2+p_x/c+\dots\ll 1$). 

Now, the maximal values of $A(t)$ and $A'(t)$ are assumed to be of the same order of magnitude which is justified by the following example.
For a typical laser pulse of sinusoidal form with an envelope $R(t)$ one finds
\begin{subequations}
\label{ClEqSin}
\begin{align}
\label{ClEqSina}
A(t)&=R(t)\cdot\sin(\t)\\
\label{ClEqSinb}
A'(t)&=\frac{\dot{R}(t)}{\omega}\cdot\sin(\t)+R(t)\cdot\cos(\t)\,.
\end{align}
\end{subequations}
A laser pulse which is not too short will reach its maximum $R_0$ only after several quarters of a period $T$, i.e. the slope of the envelope will typically fulfill the condition $\dot{R}(t)\ll R_0/(T/4)\approx\omega R_0$. With that it follows from (\ref{ClEqSin}) that the maximal values of $A(t)$ and $A'(t)$ are of the same order of magnitude.

In the following, only those terms are kept that are not higher than second order in the small quanities $A/c^2$, $A'/c^2$, $p_x/c$ and $kz$. Thus the integral in (\ref{ClEq4}) vanishes and the equation can be integrated to give an explicite solution for the motion in z-direction:
\begin{equation}
\label{ClEq5}
z=\frac{1}{2c^3}\int A^2 dt-\frac{p_x}{c^2}\int A dt+p_z\cdot t+z_0\,.
\end{equation} 

This can be inserted into (\ref{ClEq3a}), and after neglecting the high order terms again, this can be integrated:
\begin{equation}
\label{ClEq6}
x=-\frac{1}{c}\left(1+\frac{p_z}{c}\right)\int A dt+\frac{1}{c^2}(p_z\cdot t+z_0)A+p_x\cdot t +x_0\,.
\end{equation}

The dynamics in y-direction which is defined by (\ref{ClEq2b}) is simply a free motion:
\begin{equation}
\label{ClEq7}
y=p_y\cdot t+y_0\,.
\end{equation}

The equations of motion found here could also be obtained as a limit of the fully relativistic solution \cite{Salamin} with the corresponding approximations.

\section{Quantum Mechanical Description}\label{QM1}
The quantum mechanical dynamics of an electron in a laser field is determined by the Schr\"odinger equation:
\begin{equation}
\label{QMDyn1}
i\frac{\partial}{\partial t}\psi=\frac{1}{2}\left[-i\vec{\nabla}-\frac{1}{c}\vec{A}\right]^2\psi=\frac{1}{2}\left[-\vec{\nabla}^2+\frac{2i}{c}(A-kz\cdot A')\frac{\partial}{\partial x}+\frac{1}{c^2}A^2-\frac{2}{c^2}AA'\right]\psi\,.
\end{equation}
Note here that the vector potential is assumed to be in the Coulomb gauge, i.e. $\vec{\nabla}\cdot\vec{A}=0$.
The particle is supposed to be localized in the vicinity of $z=0$, i.e. terms of order $(kz)^2$ are neglected. This will also be employed to find the solution of the Schr\"odinger equation. It will be solved separately for the orders $(kz)^0$ and $(kz)^1$ using the following ansatz:
\begin{equation}
\label{QMDyn2}
\psi=\exp i\big[p_x(x-x_0)+p_y(y-y_0)+p_z(z-z_0)+u(t)\cdot kz+w(t)\big]\,.
\end{equation}
With that the partial differential equation reduces to two ordinary differential equations, one for each order of $(kz)$:
\begin{subequations}
\label{QMDyn3}
\begin{align}
\dot{u}&=-\frac{p_x}{c}A'+\frac{1}{c^2}AA'\\
\dot{w}&=-\frac{1}{2}\Big[p_x^2+p_y^2+p_z^2+2kp_z\cdot u +k^2u^2-\frac{2p_x}{c}\cdot A+\frac{1}{c^2}A^2\big]\,.
\end{align}
\end{subequations}
These equations can be solved by mere integration:
\begin{subequations}
\label{QMDyn4}
\begin{align}
u=&-\frac{p_x}{\omega c}A+\frac{1}{2\omega c^2}A^2\\
w=&-\frac{1}{2}\big[p_x^2+p_y^2+p_z^2\big]t+\frac{p_x}{c}\left(1+\frac{p_z}{c}\right)\int A dt\\\nonumber
&-\frac{1}{2c^2}\left(1+\frac{p_z}{c}+\frac{p_x^2}{c^2}\right)\int A^2 dt+\frac{p_x}{2c^5}\int A^3 dt-\frac{1}{8c^6}\int A^4 dt\,.
\end{align}
\end{subequations}
Now, those terms will be dropped that are by a second order factor of $A/c^2$ less than other terms.
Thus the solution of the Schr\"odinger equation reads
\begin{multline}
\label{QMDyn5}
\psi=\exp i\left[p_x(x-x_0)+p_y(y-y_0)+p_z(z-z_0)+\left(-\frac{p_x}{\omega c}A+\frac{1}{2\omega c^2}A^2\right)\cdot kz\right.\\
\left.-\frac{1}{2}\big[p_x^2+p_y^2+p_z^2\big]t+\frac{p_x}{c}\left(1+\frac{p_z}{c}\right)\int A dt
-\frac{1}{2c^2}\left(1+\frac{p_z}{c}\right)\int A^2 dt\right]\,.
\end{multline}
This wave function was found under the assumption that the particle be localized in z-direction, i.e. a wave-packet has to be constructed out of these non-localized wave functions to get a consistent solution. That will be subject of the next chapters.

\section{Gaussian Wave Packets}
A convenient method is to work with gaussian wave packets. They possess well-defined widths in both momentum and coordinate space, and they are rather easy to handle.

Now, functions of the general form
\begin{equation}
\label{GaWP1}
\phi=\frac{1}{\sqrt{2\pi}}\exp\big[if(x,p,t)\big]
\end{equation}
are being superimposed with a gaussian shaped weighting factor
\begin{equation}
\label{GaWP2}
\xi=\frac{1}{\sqrt{\Delta p\sqrt{\pi/2}}}\cdot\exp-\frac{(p-p_0)^2}{\Delta p^2}
\end{equation}
to define the wave function $\Psi$:
\begin{equation}
\label{GaWP3}
\Psi(x,t)=\big(\pi\sqrt{2\pi}\Delta p\big)^{-1/2}\int^\infty_{-\infty}\exp-\frac{(p-p_0)^2}{\Delta p^2}\cdot\exp\big[if(x,p,t)\big]dp\,.
\end{equation}
Here $f(x,p,t)$ is a general real function that depends on time, one momentum and one spacial variable. The maximum of the momentum distribution is $p_0$ and its width is given by $\Delta p$.
If both functions are normalized according to
\begin{subequations}
\label{GaWP4}
\begin{align}
\int^\infty_{-\infty}\phi^*(x,p,t)\phi(x,\tilde{p},t) dx&=\delta(p-\tilde{p})\\
\int^\infty_{-\infty}\xi^*(p)\xi(p)dp&=1\,.
\end{align}
\end{subequations}
then this holds for $\Psi$ as well:
\begin{equation}
\label{GaWP5}
\int^\infty_{-\infty}\Psi^*(x)\Psi(x)dx=\int^\infty_{-\infty}
\phi^*(x,p,t)\xi^*(p)\phi(x,\tilde{p},t)\xi(\tilde{p})\,
dp\,d\tilde{p}\,dx=1\,.
\end{equation}
Now the function $f(x,p,t)$ is expanded with respect to $p$ at the point $p_0$:
\begin{equation}
\label{GaWP6}
f(p)=f(p_0)+f'(p_0)(p-p_0)+\smfrac{1}{2}f''(p_0)(p-p_0)^2\,.
\end{equation}
This can be a good approximation if the momentum $p$ is close to $p_0$ due to a narrow width $\Delta p$ in momentum space. If $f(x,p,t)$ is only quadratic in $p$, as will be the case here, the expansion will even be exact. Now the integral (\ref{GaWP3}) can be done by means of
\begin{equation}
\label{GaWP7}
\int^\infty_{-\infty}\exp(-\alpha x^2+\beta x)dx=\sqrt{\frac{\pi}{\alpha}}\exp{\frac{\beta^2}{4\alpha}}\,.
\end{equation}
The wave function $\Psi(x)$ and the probability density $|\Psi(x)|^2$ are found to be
\begin{subequations}
\label{GaWP8}
\begin{align}
\label{GaWP8a}
&\Psi(x)=\frac{\sqrt{\Delta p}}{\sqrt[4]{2\pi}}\left(1-\frac{i}{2}\Delta p^2f''(p_0)\right)^{-1/2}\exp\big(if(p_0)\big)\cdot\exp-\frac{\left[\frac{\Delta p}{2}f'(p_0)\right]^2}{1-\frac{i}{2}\Delta p^2f''(p_0)}\\
\label{GaWP8b}
|&\Psi(x)|^2=\frac{\Delta p}{\sqrt{2\pi}}\left(1+\big(\smfrac{\Delta p^2}{2}f''(p_0)\big)^2\right)^{-1/2}\exp-\frac{\frac{\Delta p^2}{2}\cdot f'(p_0)^2}{1+\big(\frac{\Delta p^2}{2}f''(p_0)\big)^2}\,.
\end{align}
\end{subequations}
It can be read off, that the maximum of the wave packet is determined by the derivative $f'(x,p,t)$ whereas the second derivative $f''(x,p,t)$ is responsible for the wave packet spread.

\section{Wave Packet Dynamics}\label{QM3}
The results of the last section can now be applied to construct a localized wave packet out of the wave functions (\ref{QMDyn5}). The formula for the probability density (\ref{GaWP8}) has been derived for one dimension. However, the three dimensional problem at hand can be reduced to three one dimensional ones, i.e. the wave function (\ref{QMDyn5}) can be rewritten in product form
\begin{equation}
\label{WPDyn1}
\psi(p_x,p_y,p_z)=\exp(\tilde{p}_x)\cdot\exp(\tilde{p}_y)\cdot\exp(\tilde{p}_z)\,.
\end{equation}
For that, the mixed term $p_x p_z$ in (\ref{QMDyn5}) needs to be removed, which is accomplished by means of a coordinate transformation. The $p_y$-dependence separates from the beginning, i.e. the transformation is only in x- and z-direction. The terms $E_q$ in the exponent (\ref{QMDyn5}) that are quadratic in $p_x$ and $p_z$ are written in the following form
\begin{equation}
\label{WPDyn2}
E_q=-\frac{1}{2}\Bigg(\begin{array}{c}p_x\\p_z\end{array}\Bigg)^{\mbox{\small T}}\,\Bigg(\begin{array}{cc}t & -\frac{1}{c^2}\int A dt\\ -\frac{1}{c^2}\int A dt & t\end{array}\Bigg)\Bigg(\begin{array}{c}p_x\\p_z\end{array}\Bigg)\,,
\end{equation}
where $T$ denotes the transpose of a matrix.
The $2\times 2$-matrix in (\ref{WPDyn2}) possesses the following eigenvectors $v_{1/2}$ with the eigenvalues $\lambda_{1/2}$:
\begin{equation}
\begin{array}{cc}
\label{WPDyn3}
\displaystyle{v_1=\frac{1}{\sqrt{2}}{1 \choose 1}}\quad & \displaystyle{\lambda_1=t-\frac{1}{c^2}\int A dt}\\[1.5em]
\displaystyle{v_2=\frac{1}{\sqrt{2}}{-1 \choose 1}}\quad & \displaystyle{\lambda_2=t+\frac{1}{c^2}\int A dt}
\end{array}
\end{equation}
From that one finds the following transformation law:
\begin{equation}
\label{WPDyn4}
\begin{aligned}
\tilde{x}=\frac{1}{\sqrt{2}}\big(x+z\big)\qquad x=\frac{1}{\sqrt{2}}\big(\tilde{x}-\tilde{z}\big)\\
\tilde{z}=\frac{1}{\sqrt{2}}\big(-x+z\big)\qquad z=\frac{1}{\sqrt{2}}\big(\tilde{x}+\tilde{z}\big)
\end{aligned}
\end{equation}
The new coordinate system is related to the old one by a rotation of $\pi/4$ to the left. In the new coordinates the wave function (\ref{QMDyn5}) can be rewritten in product form:
\begin{equation}
\label{WPDyn5}
\begin{aligned}
\psi&={\textstyle\exp i\left[p_y(y-y_0)-\frac{1}{2}p_y^2 t-\frac{1}{2c^2}\int A^2 dt\right]}\\
&{\textstyle\times
\exp i\left[\tilde{p}_x\left(\tilde{x}-\tilde{x}_0-\frac{1}{2c^2}A(\tilde{x}+\tilde{z})+\frac{1}{\sqrt{2}c}\int A dt-\frac{1}{2\sqrt{2}c^3}\int A^2 dt\right)-\frac{1}{2}\tilde{p}_x^2\left(t-\frac{1}{c^2}\int A dt\right)\right]}\\
&{\textstyle\times
\exp i\left[\tilde{p}_z\left(\tilde{z}-\tilde{z}_0+\frac{1}{2c^2}A(\tilde{x}+\tilde{z})-\frac{1}{\sqrt{2}c}\int A dt-\frac{1}{2\sqrt{2}c^3}\int A^2 dt\right)-\frac{1}{2}\tilde{p}_z^2\left(t+\frac{1}{2c^2}\int A dt\right)\right]}
\end{aligned}
\end{equation} 
Applying (\ref{GaWP8b}) to this wave function one finds the following expression for the probability density:
\begin{equation}
\label{WPDyn6}
\begin{aligned}
|&\psi(\tilde{x},\tilde{y},z)|^2=\frac{\Delta p_y}{\sqrt{2\pi}}\left[1+\big(\smfrac{\Delta p_y^2}{2}t\big)^2\right]^{-1/2}\exp-\frac{\frac{\Delta p_y^2}{2}\big[y-y_0-p_y t\big]^2}{1+\big(\frac{\Delta p_y^2}{2}t\big)^2}\\
&\times\frac{\Delta\tilde{p}_x}{\sqrt{2\pi}}\left[1+\left(\smfrac{\Delta\tilde{p}_x^2}{2}{\textstyle\left(t-\frac{1}{c^2}\int A dt\right)}\right)^2\right]^{-1/2}
\cdot\frac{\Delta\tilde{p}_z}{\sqrt{2\pi}}\left[1+\left(\smfrac{\Delta\tilde{p}_z^2}{2}{\textstyle\left(t+\frac{1}{c^2}\int A dt\right)}\right)^2\right]^{-1/2}\\
&\times
\exp-\frac{\frac{\Delta\tilde{p}_x^2}{2}\left[
\tilde{x}-\tilde{x}_0-\frac{1}{2c^2}A(\tilde{x}+\tilde{z})+\frac{1}{\sqrt{2}c}\int A dt-\frac{1}{2\sqrt{2}c^3}\int A^2 dt
-\tilde{p}_x\left(t-\frac{1}{c^2}\int A dt\right)
\right]^2}{1+\left(\frac{\Delta\tilde{p}_x^2}{2}\left(t-\frac{1}{c^2}\int A dt\right)\right)^2}\\
&\times
\exp-\frac{\frac{\Delta\tilde{p}_z^2}{2}\left[
\tilde{z}-\tilde{z}_0+\frac{1}{2c^2}A(\tilde{x}+\tilde{z})-\frac{1}{\sqrt{2}c}\int A dt-\frac{1}{2\sqrt{2}c^3}\int A^2 dt
-\tilde{p}_z\left(t+\frac{1}{c^2}\int A dt\right)
\right]^2}{1+\left(\frac{\Delta\tilde{p}_z^2}{2}\left(t+\frac{1}{c^2}\int A dt\right)\right)^2}
\end{aligned}
\end{equation}
The motion of the maximum $\vec{x}_m (t)$ of this gaussian wave packet is easily found by equating the arguments of the exponential functions to zero. It is instructive to transfer these equations back to the original coordinates by means of (\ref{WPDyn4}). The following equations are found:
\begin{subequations}
\label{WPDyn7}
\begin{align}
x_m&=-\frac{1}{c}\left(1+\frac{p_z}{c}\right)\int A dt+\frac{1}{c^2}\cdot z_m\cdot A+p_x\cdot t+x_0\\\nonumber
&=-\frac{1}{c}\left(1+\frac{p_z}{c}\right)\int A dt+\frac{1}{c^2}(p_z\cdot t+z_0)\cdot A+p_x\cdot t+x_0\\
z_m&=\frac{1}{2c^3}\int A^2 dt-\frac{p_x}{c^2}\int A dt+p_z\cdot t +z_0\\
y_m&=p_y\cdot t+y_0\,.
\end{align}
\end{subequations}
These are identical to the equations of motion of a classical particle (\ref{ClEq5},\ref{ClEq6},\ref{ClEq7}). 

To fully understand the wave packet dynamics it is helpful to examine the directions of the principal axes (i.e. the directions of those axes where the widths of the wave packet are extremal) and the widths of the gaussians along these axes. 
The probability density (\ref{WPDyn6}) relative to its maximum is proportional to
\begin{equation}
\label{WPDyn8}
|\psi|^2\propto\exp-\left[\frac{\frac{\Delta p_y^2}{2}\cdot y^2}{1+\big(\frac{\Delta p_y^2}{2}t\big)^2}
+\frac{\frac{\Delta\tilde{p}_x^2}{2}\left[
\tilde{x}-\frac{1}{2c^2}A(\tilde{x}+\tilde{z})
\right]^2}{1+\left(\frac{\Delta\tilde{p}_x^2}{2}\left(t-\frac{1}{c^2}\int A dt\right)\right)^2}
+\frac{\frac{\Delta\tilde{p}_z^2}{2}\left[
\tilde{z}+\frac{1}{2c^2}A(\tilde{x}+\tilde{z})
\right]^2}{1+\left(\frac{\Delta\tilde{p}_z^2}{2}\left(t+\frac{1}{c^2}\int A dt\right)\right)^2}\right]\,.
\end{equation}
The coordinate transformation needed to find the normal form of the exponent will only concern the x- and the z-direction.
Although the exact transformation can be found the result is complicated and not very enlightening. It is much more instructive and less cumbersome to consider an approximative transformation for the case that the initial wave packet is axially symmetric (with respect to the y-axis), i.e. $\Delta\tilde{p}_x=\Delta\tilde{p}_z\equiv\Delta\tilde{p}$. Furthermore, the terms $1/c^2\int A dt$ are considered to be small compared to $t$. This is a reasonable assumption for two reasons. Firstly, the dynamics is non-relativistic, i.e. $(A/c^2)^2\ll 1$ (see section \ref{ClEq}) and secondly, the integral partly cancels if the vector potential describes an oscillating laser pulse.
With that, the following quadratic form $QF$ needs to be transformed to its normal form:
\begin{equation}
\label{WPDynQF}
QF=
\frac{\frac{\Delta\tilde{p}^2}{2}}
{1+\left(\frac{\Delta\tilde{p}^2}{2}t\right)^2}
\left[\left(\tilde{x}-\frac{1}{2c^2}A(\tilde{x}+\tilde{z})\right)^2
+\left(\tilde{z}+\frac{1}{2c^2}A(\tilde{x}+\tilde{z})\right)^2\right]\,.
\end{equation}
 Now, the following orthogonal transformation is applied:
\begin{equation}
\label{WPDyn9}
\Bigg(\begin{array}{c}X\\Z\end{array}\Bigg)=
\Bigg(\begin{array}{cc} 1 & \frac{1}{4c^2}A\\ -\frac{1}{4c^2}A & 1\end{array}\Bigg)
\Bigg(\begin{array}{c}\tilde{x}\\\tilde{z}\end{array}\Bigg)
\hspace{1.5cm}
\Bigg(\begin{array}{c}\tilde{x}\\\tilde{z}\end{array}\Bigg)=
\Bigg(\begin{array}{cc} 1 & -\frac{1}{4c^2}A\\ \frac{1}{4c^2}A & 1\end{array}\Bigg)
\Bigg(\begin{array}{c}X\\Z\end{array}\Bigg)
\end{equation}
To the order $(A/c^2)^1$ this is a rotation of the coordinate system by an angle of $A(t)/(4c^2)$ to the left. With $(A/c^2)^2\ll 1$, the quadratic form (\ref{WPDynQF}) is diagonalized and the probability density becomes
\begin{equation}
\label{WPDyn10}
|\psi|^2\propto\exp-\left[\frac{\frac{\Delta p_y^2}{2}}{1+\big(\frac{\Delta p_y^2}{2}t\big)^2}\cdot y^2
+\frac{\frac{\Delta\tilde{p}^2}{2}}
{1+\left(\frac{\Delta\tilde{p}^2}{2}t\right)^2}
\left[\left(1-\frac{1}{c^2}A\right)X^2
+\left(1+\frac{1}{c^2}A\right)Z^2\right]
\right]\,.
\end{equation}
The corresponding widths (for the definition see (\ref{GaWP2})) are given by
\begin{subequations}
\label{WPDyn11}
\begin{align}
\Delta X^2&=\frac{2}{\Delta\tilde{p}^2}\left[\left(1+\frac{1}{2c^2}A\right)\left(1+\left(\frac{\Delta\tilde{p}^2}{2}t\right)^2\right)\right]\\
\Delta Z^2&=\frac{2}{\Delta\tilde{p}^2}\left[\left(1-\frac{1}{2c^2}A\right)\left(1+\left(\frac{\Delta\tilde{p}^2}{2}t\right)^2\right)\right]\\
\Delta y^2&=\frac{2}{\Delta p_y^2}\left(1+\left(\frac{\Delta p_y^2}{2}t\right)^2\right)\,.
\end{align}
\end{subequations}
With that the ratio of $\Delta X$ and $\Delta Z$ is found to be
\begin{equation}
\label{WPDyn12}
\frac{\Delta X}{\Delta Z}=1+\frac{1}{2c^2}A\,.
\end{equation}

\section{Discussion}
The equations of motion for the maximum of the laser-driven wave packet (\ref{WPDyn7}), which are identical to the classical motion, contain several terms that are proportional to the vector potential $A(t)$ or its integral $\int A(t) dt$. Typically, these are terms that oscillate with the laser frequency $\omega$, such as the term responsible for the motion in E-field direction (-1/c $\int A(t) dt$).
A different term $\int A(t)^2 dt$ is contained in the equation for the z-direction, which is responsible for the particle drift in laser-propagation direction. Since the integral is over the square of the vector potential this term grows as the laser pulse passes, no matter what shape it has. That means, a particle will always be displaced in propagation direction of a laser pulse. This is understood by the magnetic Lorentz force that is excerted on the particle as it moves in polarization direction.
 It is necessary to go beyond the dipole approximation to find this term.
Finally, there is a constant motion term $(p\cdot t+const)$ for each direction that is determined by the initial conditions.  

The wave packet itself is characterized by its principle axes. Their directions are those of the basis vectors of the initial coordinate system which is turned by an angle of $\pi/4+A(t)/(4c^2)$ about the y-axis. The widths of the gaussians along the principle axes are given by (\ref{WPDyn11}). The terms containing the vector potential are only (typically oscillatory) corrections to the wave packet spread $\Delta x$ of a free particle
\begin{equation}
\label{Dis1}
\Delta x^2=\frac{2}{\Delta p^2}\left(1+\left(\frac{\Delta p^2}{2}t\right)^2\right)\,,
\end{equation}
which is determined by the initial momentum width $\Delta p$.
Thus the wave packet shows the well-known behavior, that the narrower the packet is initially, the faster it spreads due to the corresponding greater width in momentum space.

\begin{figure}[h]
\includegraphics{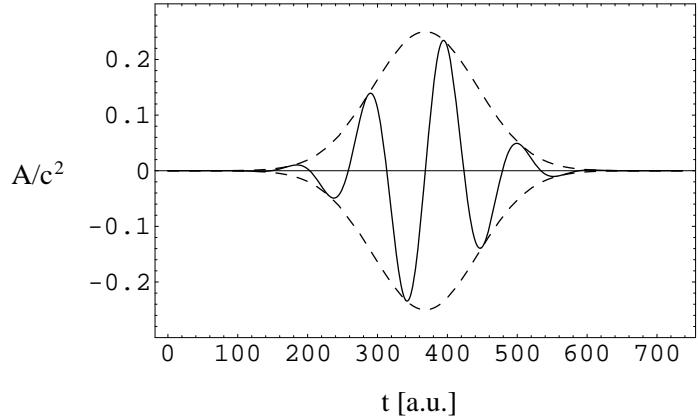}
\caption{The vector-potential in the dimensionless form $A/c^2$, which represents the laser pulse shape.
Except for a factor of four, this curve also shows the time-dependent angle between the principle axes of the electron wave packet (in the x-z-plane) and the diagonals of the coordinate system ($x=\pm z$). The angle is given by $A/(4c^2)$.}
\label{pic1}
\end{figure}

In the following example a gaussian shaped laser pulse of a few cycles is considered \mbox{(FIG. \ref{pic1})}. It is given by
\begin{equation}
\label{Dis2}
A(t)=R c^2 \sin(\t-\varphi_0)\exp\left[-\left(\frac{\t-\varphi_0}{\Delta z}\right)^2\right]\,,\quad R=0.25\,,\quad\Delta z=6\,,\quad\varphi_0=3.5\cdot\Delta z
\end{equation}
The total area of that anti-symmetric pulse is zero, i.e. $\int_{-\infty}^{\infty} A dt=0$. Thus according to (\ref{ClEq5},\ref{ClEq6},\ref{ClEq7}) the position of the particle after the pulse has passed is given by
\begin{subequations}
\label{Dis3}
\begin{align}
x&=p_x\cdot t+x_0\\
z&=\frac{1}{2c^3}\int A^2 dt+p_z\cdot t +z_0\approx 282.4+p_z\cdot t +z_0\\
y&=p_y\cdot t+y_0\,,
\end{align}
\end{subequations}
i.e. it is travelling at its initial velocity and it is displaced by roughly 280 a.u. in z-direction. The electron does not pick up any momentum,
because the final and initial values of the vector potential are identical. 
For the same reason, the widths of the wave packet (\ref{WPDyn11}) reduce to the form of a freely travelling particle (\ref{Dis1}).

\begin{figure}[h]
\includegraphics{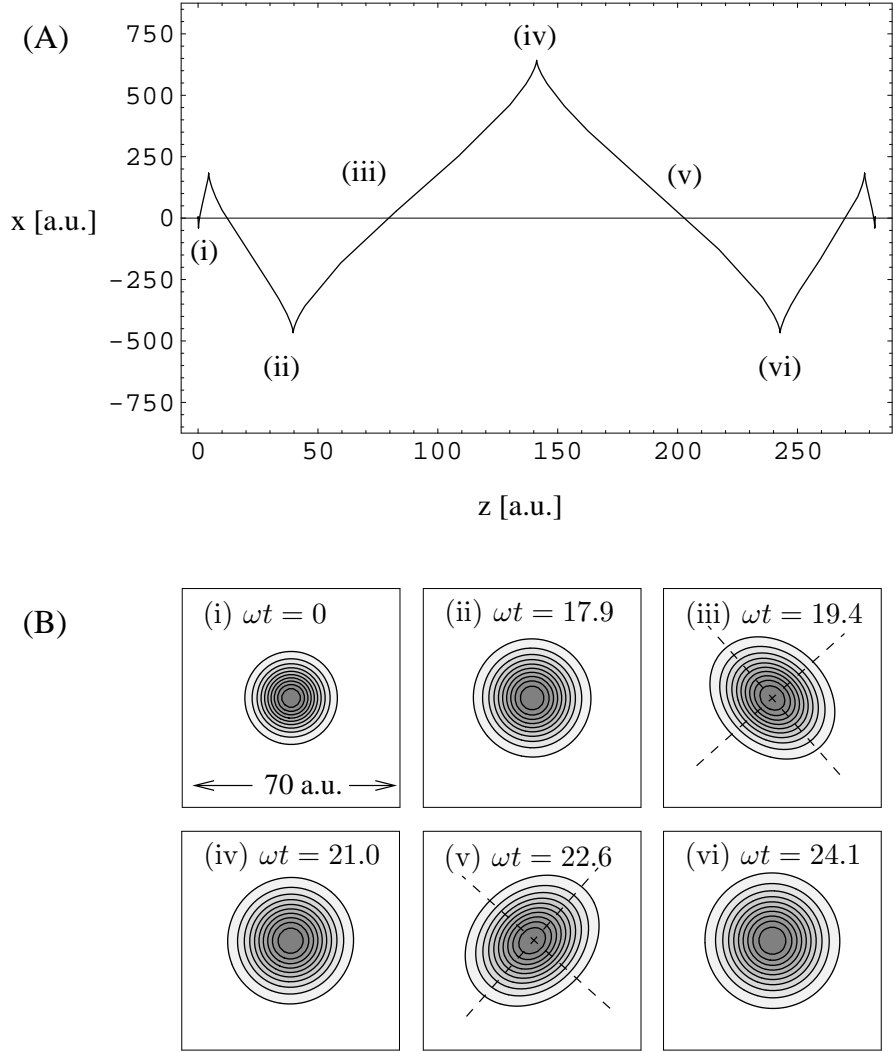}
\caption{Electron wave packet dynamics:
(A) shows the trajectory of the maximum of the wave packet. The labels (i) to (vi) mark the positions for which the wave packets are plotted in figure (B). The sides of
each of the squares have a width of $70\;a.u.$
In the pictures (iii) and (v), the principle axes of the wave packets are shown. The directions deviate from the diagonals by an angle of $A(t_{iii/v})/(4c^2)$. The other wave packets in (B) are axially symmetric.}
\label{pic2}
\end{figure}

The dynamics of the particle in x- and z-direction is depicted in (FIG. \ref{pic2}). (A) shows the motion of the maximum of the wave packet. Now, the initial conditions are chosen such that the particle rests at the origin at the beginning.
The widths in momentum space are $\Delta p_x=\Delta p_z=0.05\;a.u.$. The wave packet is shown in (B) for the times when the maximum is located at the positions marked in (A). These correspond to the times $t_i=0$, $\t_{ii/vi}-\varphi_0=\pm\pi$, $\t_{iii/v}-\varphi_0=\pm\pi/2$ and $\t_{iv}=\varphi_0$, where $\varphi_0$ determines the maximum of the envelope (see (\ref{Dis2})).
It is seen that the wave packet, while spreading, deforms according to (\ref{WPDyn12}).

We note finally that pure quantum effects due to self interferences with negative Wigner functions are possible for the described scenario \cite{Mahmoudi}.

\begin{thebibliography}{02}
\bibitem{Volkov}D.M. Volkov, Z. Phys. {\bf 94}, 250 (1935).
\bibitem{Joachain}N.J. Kylstra, A.M. Ermolaev, and C.J. Joachain, J. Phys. B {\bf30}, 449 (1997)
\bibitem{Knight}U.W. Rathe, C.H. Keitel, M. Protopapas, and P.L. Knight, J. Phys. B {\bf30}, 531 (1997)
\bibitem{Grobe}J.W. Braun, Q. Su, and R. Grobe, Phys. Rev. A {\bf 59}, 604 (1998).
\bibitem{Mocken}G. Mocken and C.H. Keitel, J. Comput. Phys. {\bf 199}, 558 (2004).
\bibitem{Roso}J.S. Roman, L. Roso, and H.R. Reiss, J. Phys. B {\bf 33}, 1869 (2000).
\bibitem{Friedrich}H. Friedrich, {\it Theoretical atomic physics; 2. rev. and enl. ed.} (Springer, Heidelberg, 1998).
\bibitem{Salamin}Y.I. Salamin and F.H. Faisal, Phys. Rev. A {\bf 54}, 4383 (1996).\bibitem{Mahmoudi}M. Mahmoudi, Y.I. Salamin, and C.H. Keitel, submitted.
\end{thebibliography}
 \end{document}